# Flexible Production Systems: Automated Generation of Operations Plans Based on ISA-95 and PDDL

Bernhard Wally, Jiří Vyskočil, Petr Novák, Christian Huemer, Radek Šindelář, Petr Kadera, Alexandra Mazak, and Manuel Wimmer

*Abstract*—Model-driven engineering (MDE) provides tools and methods for the manipulation of formal models. In this letter, we leverage MDE for the transformation of production system models into flat files that are understood by general purpose planning tools and that enable the computation of "plans", i.e., sequences of production steps that are required to reach certain production goals. These plans are then merged back into the production system model, thus enriching the formalized production system knowledge.

*Index Terms*—AI-based methods, factory automation, intelligent and flexible manufacturing.

## I. INTRODUCTION

MANUFACTURING systems of the future are required to be more and more flexible, regarding both the products they produce and the production systems themselves [1], [2]. According to the principles of smart manufacturing, products and their recipes are not required to be known at design time, product variants may be edited at runtime, and production planning and scheduling are to be invoked on-the-fly, when a new production order appears. As such, the use of automated planning systems seems very natural, however, current commercial industrial planning systems are not sufficient [3].

Processing production orders on-the-fly means that a flexible manufacturing line does not need to be in a predefined initial state before starting a new production. Moreover, the manufacturing line can even be already producing other orders, and thus the state of all resources such as shuttles on a transportation system, or locations of material can vary. Moving these shuttles back to an artificial initial state, as it is done in industrial practice currently, would mean time and energy loss that could and should be avoided. Such high degrees of freedom disqualify traditional ways of programming manufacturing lines and strengthen the need for using automated planning systems being able to react on changing initial conditions and targets. Further, a declarative way of programming related to planners and industrial specification languages is essential for fulfilling the challenging demands of smart manufacturing systems.

In this letter, we are presenting a model-driven approach to automatically transform a manufacturing system specification to a production plan via automated planning. To formulate the manufacturing line planning task, a specification of all industrial components and their actions and interactions is needed. In this environment a number of methods, tools and standards are well established:

1) *Production systems engineering*: specification of industrial components and processes using industry standards or domain specific modeling languages [4], [5];
2) *Model-driven engineering* (MDE): generic methods for the specification of discrete models, their validation, manipulation and transformation, etc. [6]–[8];
3) *Automated reasoning*: methods for realizing reasoning tasks, covering various classes of problems with different computational complexities, from NP-complete propositional logic, EXPSPACE-complete classical planning, semi-decidable first-order logic, up to undecidable halting problems [9]–[11].

Our proposed approach covers both the engineering phase of systems as well as their runtime. With respect to the engineering phase, it can be considered as a verification tool for the fulfillment of functional requirements, with respect to the runtime it can make use of automated reasoning in order to find sequences of production steps to reach initially unknown production system states or to produce products that have not been known at design time. The overview of the proposed approach is depicted in Fig. 1.

This contribution is structured as follows: after a discussion of related work in Section II, Section III describes the rules for transforming production system models into planning problems, while Section IV discusses their application on a specific use case. Section V presents concrete problem statements as well as

Manuscript received February 15, 2019; accepted June 27, 2019. Date of publication July 22, 2019; date of current version August 8, 2019. This letter was recommended for publication by Associate Editor A. Agostini and Editor T. Asfour upon evaluation of the reviewers' comments. This letter was supported in part by the Austrian Federal Ministry for Digital and Economic Affairs and the Austrian National Foundation for Research, Technology and Development, in part by the DAMiAS project funded by the Technology Agency of the Czech Republic, in part by the H2020 project DIGICOR, and in part by the OP VVV DMS project Cluster 4.0. *(Corresponding author: Bernhard Wally.)*

B. Wally, R. Šindelář, A. Mazak, and M. Wimmer are with the CDL for Model-Integrated Smart Production, Department of Business Informatics – Software Engineering, Johannes Kepler University Linz, Linz 4040, Austria (e-mail: bernhard.wally@jku.at; radek.sindelar@jku.at; alexandra.mazak@jku.at; manuel.wimmer@jku.at).

J. Vyskočil, P. Novák, and P. Kadera are with the Department of Intelligent Systems for Industry and Smart Distribution Networks, Czech Institute of Informatics, Robotics, and Cybernetics, Czech Technical University in Prague, Prague 160 00, Czech Republic (e-mail: jiri.vyskocil@cvut.cz; petr.novak@cvut.cz; petr.kadera@cvut.cz).

C. Huemer is with the Business Informatics Group, Institute of Information Systems Engineering, Technische Universitat Wien, Vienna 1040, Austria (e-mail: huemer@big.tuwien.ac.at).

Digital Object Identifier 10.1109/LRA.2019.2929991

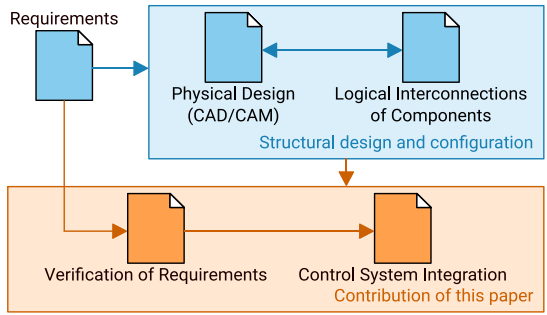

Fig. 1. High-level view on the proposed approach.

performance data of our approach. Finally, Section VI concludes and provides hints for future research directions.

## II. RELATED WORK

### A. Model-Driven Engineering

Model-driven engineering (MDE) has long been investigated and practiced. In the early 2000s, standardization efforts finally culminated in widely adopted standards such as the Meta Object Facility and the Unified Modeling Language (UML) [6]. Similar to the object-oriented paradigm of the 1980s ("everything is an object") the new paradigm was "everything is a model" [7]. Based on this foundation, MDE tools may impose domain-specific constraints and perform model checking that can detect and prevent many errors early in the life cycle [8]. This is exactly the main reason why we employ MDE techniques in this letter. We aim to formally provide knowledge of a production system and use this knowledge to verify certain properties thereof.

MDE in the context of smart production is, of course, not new. For instance, IEC 62264 (also known as ISA-95) is a series of international standards describing data structures, activities and a communication protocol in the field of manufacturing execution systems (MES) and their interfaces with enterprise resource planning (ERP) systems [4]. Specifically, parts 2 and 4 of of ISA-95 define a set of UML-based metamodels that enable the modeling of MES related information [12], [13]. With AutomationML[1] a standardized data format was introduced for representing engineering information in the area of process automation and control [5]. An integration layer for ISA-95 and AutomationML [14] has been presented in [15] and [16], enabling AutomationML to act as a container format for encoding ISA-95 information.

Model-driven transformation of transportation system knowledge from the proprietary tool PX5 Configurator has been discussed in [17], where it was converted into AutomationML before being further processed within another proprietary simulation tool. While we are also using the PX5 Configurator in this letter in our case study, we are not implementing a toolchain to achieve integration between two proprietary tools, but we define generic transformation rules between standardized (modeling) languages.

Model-driven alignment of structural production system information was further presented in [18]–[20], where business models were aligned to MES models. This letter could be an interesting extension to our work, when it comes to the integration of business information. A compatible ERP-like system has been presented in [21].

### B. Automated Planning

Automated planning is a branch of artificial intelligence that deals with the issue of finding *plans*, which are strategies or sequences of actions. Typical application scenarios are, e.g., plans that are executed by autonomous robots [22]. "Classical" planning [23] describes automated planning where a set of assumptions and restrictions have to hold.

The worst case complexity of classical planning is EXPSPACE-complete, for plan existence problems [24]. On one hand, many planning systems allow to relax some of the "classical" properties that can even lead to semi-decidability [24] of plan existence problems. On the other hand, many well-known planning problems are typically much easier (NP-complete or even better) [25].

The Planning Domain Definition Language (PDDL) is a standardized classical planning language that has been used for approximately 20 years at the international planning competitions.[2] In PDDL, the planning problem is divided into two parts: (i) the *domain* part holds all available predicates as well as the allowed actions on the state-space with preconditions and effects and (ii) one or more related *problem* part, which define the initial state and the goal state conditions. PDDL solvers then try to find sequences of actions that lead from the initial state to the goal state.

Only some of the automated reasoning methods can be utilized for (semi-)automated solutions at industrial scale. Therefore, we focused on PDDL-based classical planning problems in this letter. The latest version of the language is PDDL 3.1 [26] but there exist many variants/extensions that support various features like goal rewards, probabilistic effects, multi-agent planning, temporal planning, etc. An overview of several extensions of PDDL including explanation of techniques in successful solvers is provided in [27].

A study on usage of PDDL for a collection of typical basic industrial problems is presented in [3]. Compared to [3], the approach proposed in this letter (i) utilizes PDDL as an intermediate format rather than a tool for direct modeling by experts, (ii) we use a real system of industrial scale, and (iii) we are focused on classical (i.e., non-temporary) planners due to the efficiency. While a "bridge between automation and AI planning" is described in [3], we are enhancing this concept by incorporating a standardized specification and its translation into automation and AI planning. Various systems for executing production plans have been proposed. Some of them are discussed in [28] that mainly addresses execution of production plans based on PDDL.

---

[1]cf. https://www.automationml.org/

[2]http://www.icaps-conference.org/index.php/Main/Competitions

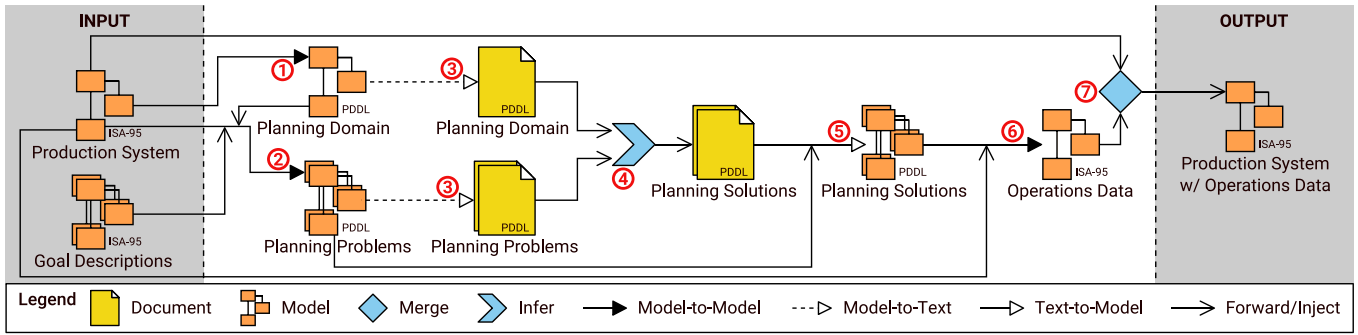

Fig. 2. Implemented core workflow of our approach. Input files need to be provided, all other artifacts are generated automatically.

## C. Synopsis

While both, model-driven engineering and automated planning, have been applied to industrial engineering, we are not aware of any particular approach which allows for automated planning solely on the model-based domain representations such as provided by ISA-95. Our proposed approach uses PDDL in a fully transparent way, i.e., the input needed by PDDL solvers is fully derived from the model and also the output provided by PDDL solvers is automatically translated back to the model. Thus, design-time and runtime decisions can be performed by domain experts without requiring knowledge about the underlying solver technology.

## III. MODEL-DRIVEN ENGINEERING OF FLEXIBLE PRODUCTION SYSTEMS

Based on the structural description of a production plant sequences of actions shall be derived that enable reaching certain production goals. We have tackled this task by leveraging (i) ISA-95 models of production systems as input and output models and (ii) PDDL as the technology for inferring sequences of actions. For this letter, the most important concepts of ISA-95 are the *equipment* and *process segment* models. Among other entities, they provides concepts for describing the machinery available in production environments such as robots, transportation systems, etc., as well as structures depicting the production steps that can be performed using the equipment. In Section III-A we describe the general approach; a detailed description follows in Section III-B, while a concrete example is presented in Section V.

## A. Approach

Our model-driven approach requires the formulation of metamodels for the involved domain models. Therefore, we have created metamodels (i) for ISA-95, following the specification given in the standards' documents and (ii) for PDDL 3.1, based on the Backus–Naur form given in [26].

We are taking two input files into account: (i) an ISA-95 model describing the production system (including equipment, material, process segments, and resource connections) and (ii) one or more ISA-95 models that describe the envisioned goal states. We will show in Section IV-A how the production system model can be automatically derived from a proprietary source model (this is an optional pre-processing step). The output is an ISA-95 model that is derived from the initial ISA-95 model, but now includes information about operations definitions. The applied "core" workflow is depicted in Fig. 2, the individual processing steps (circled numbers) are described below, accordingly.

1) **Production system → Planning domain:** the production system is parsed and relevant information extracted and transformed into PDDL domain concepts.
2) **(Production system + Goal descriptions) → Planning problem:** the production system is parsed and relevant information extracted and transformed into the initial state of a PDDL problem. For each goal description that is provided, a separate planning problem is created, with the corresponding goal specifications. The initial state of these planning problems is reused from the initially created PDDL problem.
3) **PDDL code generation:** so far, the planning domain and problems have been described by means of models. In this step, the models are serialized as plain-text PDDL documents that can be read by standard-conforming PDDL solvers.
4) **PDDL solving:** for each planning problem a planning solution is calculated by a PDDL solver. If no solution could be found for certain problems, this is also recorded. The solutions are created as plain-text files.
5) **Planning solutions → Planning solution models:** the plain-text files are "reverse-engineered" into formal PDDL models in order to be useable in the subsequent processing steps.
6) **PDDL solution models → Operations data:** the sequence of actions found by the solver is transformed into operations that are collected in an ISA-95 model.
7) **(Production system + Operations data) → Integrated model:** the original production system model and the operations data model are merged into a single ISA-95 model containing both the static production system information and the behavioral information of goal-oriented production steps.

Since step 6 generates operations data, it might be desired to generate this data at runtime instead of at design time, in order to enable flexible production systems that are able to compute production plans online. Fortunately, our approach can be applied at design time and at runtime.

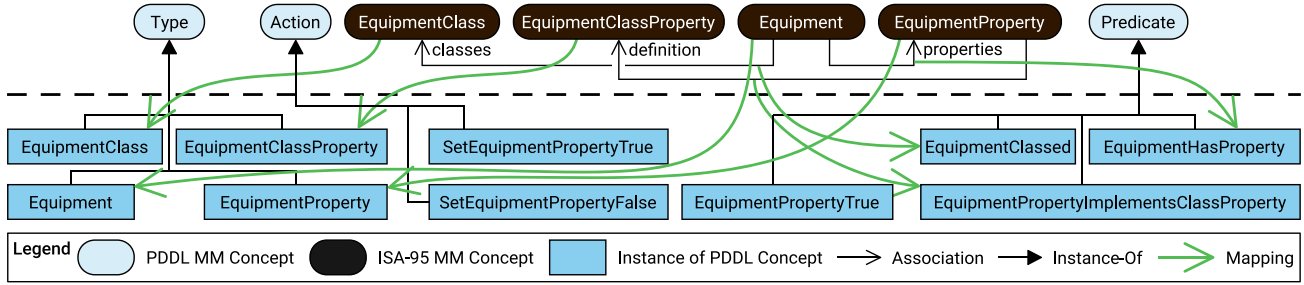

Fig. 3. Mapping of ISA-95 metamodel elements to a PDDL domain description.

## B. Implementation

We have implemented the workflow previously described based on metamodels of ISA-95 and PDDL that have been formalized using Ecore/EMF.[3] However, our approach could have been realized using any capable technology, including, e.g., ontologies. The transformation of the initial production system information into a planning domain model has been realized threefold, as described in the following two sub-sections for generic information and in Section IV-B for domain-specific concepts. Our approach assumes that ISA-95 ProcessSegments are defined in a way that they refer to EquipmentClasses rather than to Equipment, and that the runtime information uses pieces of Equipment rather than EquipmentClasses. This is typically the case.

*1) Metamodel Concepts:* relevant metamodel concepts of ISA-95 are converted to certain PDDL statements (cf. Fig. 3). (i) relevant metamodel classes (that are used by the ISA-95 model under observation) are implemented as PDDL Types. (ii) ISA-95 associations are converted to PDDL Predicates. (iii) boolean properties are supported by a dedicated Predicate, e.g., `EquipmentPropertyTrue` for equipment properties. (iv) for the manipulation of these properties, two Actions are defined: `SetEquipmentPropertyTrue` and `SetEquipmentPropertyFalse`. These two actions enable explicit setting of boolean equipment properties that are not tagged with the term `pddl:implicit` in their description attribute. A PDDL encoding of these transformed concepts is given in Lst. 1. Lines 15–16 and 21–22 encode information that takes into account instance data: properties that are tagged as being set implicitly must not be supported by the generic `SetEquipmentProperty*` actions.

It is important to note, that this is only one way of encoding an ISA-95 model in PDDL. For instance, boolean properties could instead be translated as specific Predicates and not as objects that are related to equipment instances via generic Predicates.

*2) Instance Data:* apart from preparing the PDDL environment with generic concept directly inferred from the ISA-95 metamodel, also instance data of the ISA-95 model has an impact on the planning domain description and requires proper mapping (cf. Fig. 4). Examples for the PDDL representation of this mapping are given in Lst. 2, the single mapping statements

[3]cf. https://www.eclipse.org/modeling/emf/

```
1  (:types EquipmentClass Equipment
2    EquipmentClassProperty EquipmentProperty)
3  (:predicates
4    (EquipmentClassed ?E - Equipment ?C - EquipmentClass)
5    (EquipmentPropertyImplementsClassProperty
6      ?EP - EquipmentProperty
7      ?ECP - EquipmentClassProperty)
8    (EquipmentHasProperty
9      ?E - Equipment ?P - EquipmentProperty)
10   (EquipmentPropertyTrue ?P - EquipmentProperty))
11 (:action SetEquipmentPropertyTrue
12   :parameters (?EP - EquipmentProperty)
13   :precondition (and
14     (not (EquipmentPropertyTrue ?EP))
15     (not (EquipmentPropertyImplementsClassProperty
16       ?EP ECP_PositioningUnitOccupied)))
17   :effect (EquipmentPropertyTrue ?EP))
18 (:action SetEquipmentPropertyFalse
19   :parameters (?EP - EquipmentProperty)
20   :precondition (and (EquipmentPropertyTrue ?EP)
21     (not (EquipmentPropertyImplementsClassProperty
22       ?EP ECP_PositioningUnitOccupied)))
23   :effect (not (EquipmentPropertyTrue ?EP)))
```

Listing 1. Excerpt of the PDDL domain description, showing types, predicates and actions that have been generated from the ISA-95 metamodel.

refer to specific lines in this listing (the instance data used refers to the example given in Section V):

(i) for each class instance (e.g., instances of EquipmentClass, EquipmentClassProperty), a Constant is created, using the instance's id with a suitable prefix as identifier (lines 2–4). (ii) ProcessSegments are implemented as PDDL Actions, using the id as name (line 7). The process's segment specifications are converted to parameters, as they represent the required resources for the process (line 8). Relevant ISA-95 relations are checked via specific Preconditions, using the Predicates defined in the metamodel mapping (line 11). Segment specification properties are checked for specific tags that need to be implemented in the ISA-95 model in order for the transformation process to behave as expected: if the `pddl:pre` or the `pddl:post` tag is detected, a corresponding Precondition or Effect is created, respectively (lines 12–16 and lines 18–22). Finally, the duration-related attributes of the ProcessSegment are interpreted as cost of the Action, uniformly converted to seconds (line 23).

Again, the presented conversion is just one example of encoding. For instance, if the properties were implemented as specialized predicates (as mentioned earlier) instead of as dedicated objects, the exists statement could be avoided and replaced by a simple predicate condition, as well as the forall statement could be replaced by a simple predicate effect statement issued on the corresponding Parameter, derived from the segment specification.

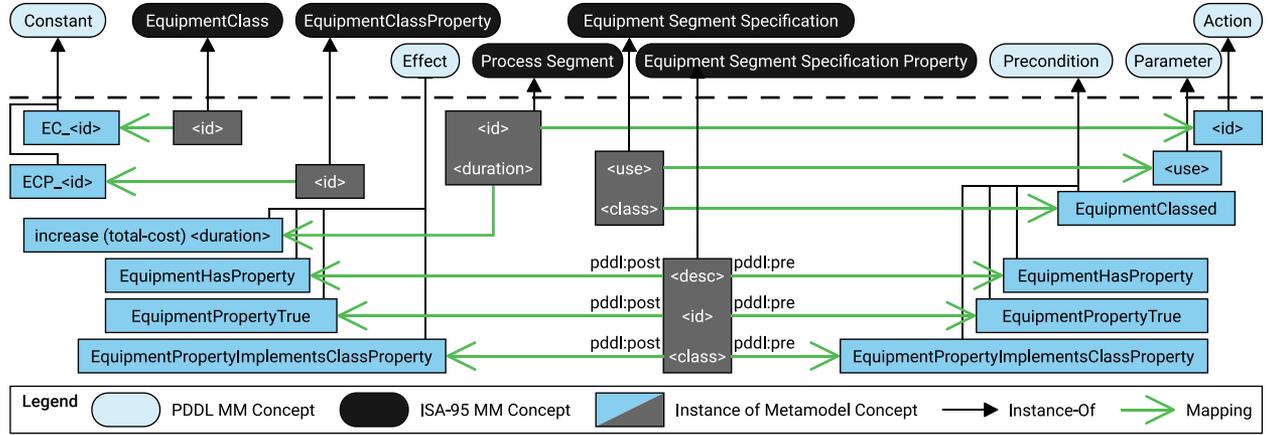

Fig. 4. Mapping of ISA-95 model elements to a PDDL domain description ("MM" means metamodel).

```
1 ;... skipping requirements and types
2 (:constants
3   EC_PositioningUnit EC_Shuttle - EquipmentClass
4   ECP_PositioningUnitOccupied - EquipmentClassProperty)
5 ;... skipping predicates
6 (:functions (total-cost))
7 (:action MoveShuttle
8   :parameters (?SHUTTLE ?FROM ?TO - Equipment)
9   :precondition (and
10    ;... skipping similar preconditions
11    (EquipmentClassed ?FROM EC_PositioningUnit)
12    (exists (?P - EquipmentProperty) (and
13      (EquipmentPropertyTrue ?P)
14      (EquipmentPropertyImplementsClassProperty
15        ?P ECP_PositioningUnitOccupied)
16      (EquipmentHasProperty ?FROM ?P))))
17   :effect (and ;... skipping similar effects
18    (forall (?P - EquipmentProperty) (when (and
19      (EquipmentPropertyImplementsClassProperty
20        ?P ECP_PositioningUnitOccupied)
21      (EquipmentHasProperty ?FROM ?P))
22      (not (EquipmentPropertyTrue ?P))))
23    (increase (total-cost) 10))
```

Listing 2. Excerpt of the PDDL domain description generated from ISA-95 instance data.

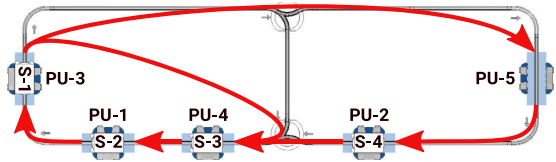

Fig. 5. Layout of the use case, rendered from within the PX5 Configurator by Montratec: positioning units PU-1 to PU-5, shuttles S-1 to S-4. Red arrows depict directed edges of the routing topology.

## IV. USE CASE – INDUSTRY 4.0 TESTBED

We are applying the mapping defined above in a use case that is derived from a real production system deployed at the Technical University in Prague, the *Industry 4.0 Testbed*. It is reduced to only the transportation system and purposely leaves out any robots or material. The physical layout of the chosen use case is depicted in Fig. 5.

Section IV-A explains how a proprietary transportation system model is attached to the workflow as an optional pre-processing step, while Section IV-B explains domain specific knowledge that is to be introduced to the core workflow.

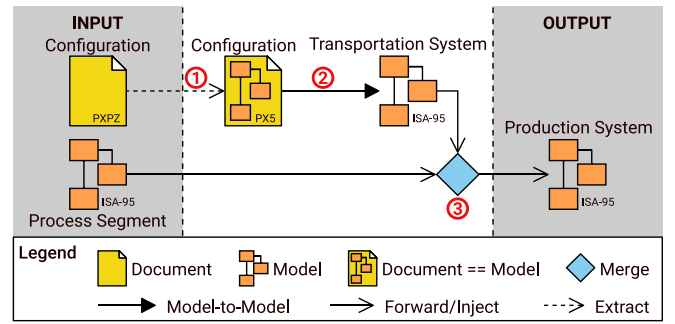

Fig. 6. Conversion workflow used to pre-process proprietary information into an ISA-95 model. The input models have been hand-crafted, the resulting output model can be used as an input model for the core workflow.

### A. PX5 Configurator

So far, the process of mapping ISA-95 elements to PDDL has been domain-agnostic. For the chosen use case of the evaluation which is situated in the field of automated intra-logistics, we need to add a few extra conversion rules in order to get meaningful results. For this, it is important to understand how the system under observation works. It is an automated transportation system centered around a monorail track that can carry one or more shuttles. These shuttles can move on the rail between so called "positioning units" (PU), which are mechatronic systems with a well-defined location on the rail that can physically lock shuttles once they are located at one of these PUs.

In order to simplify the development of a corresponding "Production System" ISA-95 model, we have implemented a converter for the proprietary tool "PX5 Configurator for montratec"; the conversion workflow is depicted in Fig. 6. In short, ① we are reading the contents of the PX5 project file, and extracting relevant information in terms of a PX5 model (a corresponding metamodel has been reverse-engineered from the underlying proprietary XML document). Then ② this PX5 model is transformed into an ISA-95 model and ③ enriched with separately modeled process information. The result of this workflow is an ISA-95 model of a production system that can be used as an input for the core workflow described in Section III and depicted in Fig. 2.

```
1  (:predicates
2    ;... skipping already defined predicates
3    (PositioningUnitConnection ?F ?T - Equipment)
4    (ShuttleLocation ?S ?PU - Equipment))
5  (:action MoveShuttle
6    ;... skipping already defined parameters
7    :precondition (and
8      ;... skipping already defined preconditions
9      (PositioningUnitConnection ?FROM ?TO)
10     (ShuttleLocation ?SHUTTLE ?FROM)
11     (not (ShuttleLocation ?SHUTTLE ?TO)))
12   :effect (and
13     ;... skipping already defined effects
14     (not (ShuttleLocation ?SHUTTLE ?FROM))
15     (ShuttleLocation ?SHUTTLE ?TO)))
```

Listing 3. Domain-specific PDDL snippets.

### B. Domain-Specific Concepts

The ISA-95 representation of this transportation system is strongly supported by the concept of ResourceRelationship-Networks. Track elements (straight line, curve and switch) are connected to each other by ResourceNetworkConnection instances of type `Track-Connection`. Positioning units and shuttles are described as "being attached" to a track element. This is realized by ResourceNetworkConnection instances of type `Positioning-Unit-Connection` and `Shuttle-Connection`, respectively, that connect these entities to corresponding track elements. Since multiple positioning units or shuttles can be located at one element, the $(x, y, z)$ coordinates of the entities are stored as FromResourceReferenceProperties. This is required for creating correct routing graphs between the PUs, as well as assigning the shuttles to the correct PUs. In the process of converting an ISA-95 model to PDDL, the track- and PU-connections are simplified to a directed graph containing only PU nodes that are connected with each other. Also, the locations of the shuttles are reduced to those of the PUs, i.e., a shuttle is only in a well-known location if it is physically located at a PU. Locations in-between are not important in the context of our planning problem. The additional mapping rules are described below; they are implemented as part of steps 1 and 2 of the *core* workflow. Line numbers below refer to Lst. 3:

i) Two Predicates are defined that correspond to the previously described simplifications: `PositioningUnit-Connection` and `ShuttleLocation`. The former allows the definition of a directed graph representing the routing scheme of the transport system (line 3). The latter describes where a certain shuttle is currently located (line 4).

ii) The ProcessSegment `MoveShuttle` defines a boolean ProcessSegmentParameter with the id `movement` and value `true`. This parameter is recognized in the first transformation step of the core workflow, from the ISA-95 model to the PDDL domain model. This process segment also specifies three EquipmentSegmentSpecifications: the shuttle `S` to move and two PUs: the source `FROM` and the destination `TO`. Based on this information, three additional Preconditions and two additional Effects are created. For the preconditions, the following statements are added: first, it is checked, whether the two positioning units are directly connected with each other (line 9). Second, it is checked whether the shuttle is currently located in the source PU (line 10). Third, it is checked if the shuttle is not already in the destination location (line 11). The last two statements are somewhat redundant in the case that all actions that manipulate the `ShuttleLocation` predicate are correctly implemented (a shuttle should never be in two places at the same time). However, this redundancy can be considered a safety net and might support comprehensiveness for human readers. The effects are clearly related: first, the shuttle is set to be no longer in the source PU (line 14) and second, the shuttle is now located in the destination PU (line 15).

## V. EVALUATION

We are evaluating our approach threefold: (i) we are evaluating the use case previously described, (ii) we are evaluating similar use cases of various sizes in order to discuss scalability aspects and (iii) we are extending these use cases to include manufacturing operations in order to exemplify transferability of the approach.

All performance numbers mentioned in the remainder have been collected on a standard portable computer, equipped with a 2.4 GHz CPU. *Fast Downward*[4] was chosen as the PDDL solver, configured with "`astar(ipdb)`".

### A. Use Case "Industry 4.0 Testbed"

Given an initial state as depicted in Fig. 5, with the four shuttles 1, 2, 3, 4 (encoded as 1234), located in PUs 3, 1, 4 and 2. The question that arose during the design time of this layout was, whether it is possible to bring the shuttles into an arbitrary order, with only one spare PU (5). While the answer to this problem might be obvious to experts, frequently, engineers are not able to answer such questions with confidence, especially if the layout is more complex. Thus, industrial systems are often equipped with additional features in an "ad-hoc" way in the hope that this would solve specific production-related problems. However, there are two problems with this approach: (i) it remains unclear whether the problem is really solved and (ii) these additional features are often quite expensive and might represent over-engineering. Therefore, an approach where the envisioned solution can be *formally verified* is clearly an advantage.

In our use case we want to find an answer to the question, and we would like to know how expensive (in terms of time) each of the reorderings is, given that each shuttle movement takes 10 s. For that, we have created 23 "goal description" ISA-95 models, each representing one of the desired goal states (excluding the goal state that is equivalent to the initial state): 1243, 1324, 1342, 1423, etc.

Lst. 4 depicts an excerpt of the generated problem definition that formulates the reordering of the shuttles from 1234 to 2341. The initial state is represented in lines 3–6, while the goal description is given in lines 8–11. Other initialization statements are left out—they are generated in step 2 of the core workflow. The 23 generated problem definitions are all exact copies of one

---

[4]cf. http://www.fast-downward.org/

```
1  (:init
2    ; other initialization left out
3    (ShuttleLocation E_Shuttle-01 E_PositioningUnit-03)
4    (ShuttleLocation E_Shuttle-02 E_PositioningUnit-01)
5    (ShuttleLocation E_Shuttle-03 E_PositioningUnit-04)
6    (ShuttleLocation E_Shuttle-04 E_PositioningUnit-02))
7  (:goal (and
8    (ShuttleLocation E_Shuttle-02 E_PositioningUnit-03)
9    (ShuttleLocation E_Shuttle-03 E_PositioningUnit-01)
10   (ShuttleLocation E_Shuttle-04 E_PositioningUnit-04)
11   (ShuttleLocation E_Shuttle-01 E_PositioningUnit-02)))
```

Listing 4. Excerpt of the PDDL init state and the complete goal description that has been generated from one of the ISA-95 goal description models.

```
1  (moveshuttle e_shuttle-01
2    e_positioningunit-03 e_positioningunit-05)
3  (moveshuttle e_shuttle-02
4    e_positioningunit-01 e_positioningunit-03)
5  (moveshuttle e_shuttle-03
6    e_positioningunit-04 e_positioningunit-01)
7  (moveshuttle e_shuttle-04
8    e_positioningunit-02 e_positioningunit-04)
9  (moveshuttle e_shuttle-01
10   e_positioningunit-05 e_positioningunit-02)
11 ; cost = 50 (general cost)
```

Listing 5. The generated plan for re-ordering the shuttles from 1-2-3-4 to 2-3-4-1. Note that the chosen solver converts all entities to lower case.

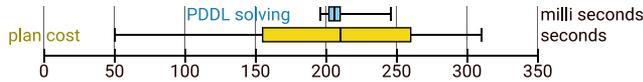

Fig. 7. Box plots of collected data from the 23 problem statements: (top) performance data of the PDDL solver, (bottom) cost of the solutions. Solving time was collected from within the Java-based workflow, i.e., it includes calling the solver as an external program, parsing its output, etc.

another, except for the goal statements that depict the desired order of the shuttles.

Lst. 5 depicts the resulting plan for the given problem. It shows that it is possible to reorder the shuttles by using five sequential steps, each costing 10s, thus resulting in a total time of 50 s, if all movements are executed sequentially.

Running the complete workflow on our concrete use case, from reading in the .pxpz file and the 23 goal statement models to the final production system model including the knowledge gained from the problem solver takes about 10 s, of which approximately 5 s are dedicated to the PDDL solver. More detailed PDDL-related data is given in Fig. 7: there, the distribution of the 23 solving times and the distribution of the 23 plan cost are depicted using box plots.

### B. Use Case "Scalability"

Scalability has been tested by generating the above mentioned use case in various sizes; not fixed to 5 PUs. We have created six instances of the transportation system, with 5 to 15 positioning units, respectively. The layout template is depicted in the lower part of Fig. 8. A load factor of 65% (ratio of the number of shuttles and the number of PUs) is implemented. The task was to find a sequence of actions that would reverse the order of the shuttles, just as in the previous use case. The results of these experiments are depicted in the upper part of Fig. 8, represented by solid lines.

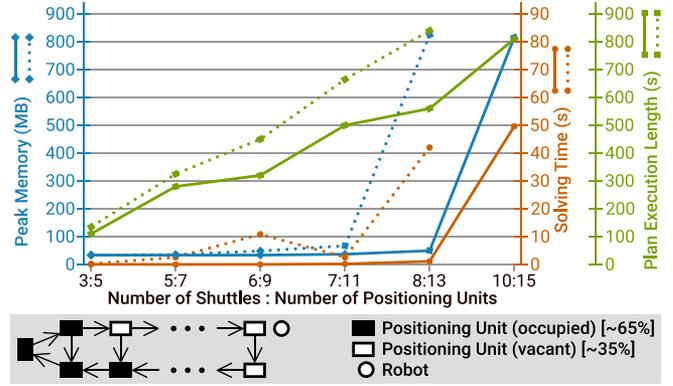

Fig. 8. Performance data gained from setups of different sizes. In these experiments, the solver has been called directly, not from within the Java-based workflow. Plan execution length is given in seconds (divide by 10 to get the number of steps required).

It can be seen that it is feasible to compute transportation plans for the given topology and load factor for settings as large as 10 shuttles on 15 PUs. This would account for small to medium sized systems. In the largest case that has been tested, computation took ≈ 50 s, which can be considered very responsive, given that the plan execution length of the corresponding solution amounts to 810 s. What can also be seen is that the length of the computed plans increases linearly, while the computational effort (memory and runtime) grows exponentially. In order to compute solutions for larger systems, it will be necessary to find either a better streamlined encoding of the ISA-95 model in PDDL, or to divide larger problems into smaller sub-problems and solving them independently from each other.

### C. Use Case "Transferability"

Production systems usually handle and alter material, which is why we have created an extended version of the use case previously described. This extended version adds material to the setting, namely wooden boards that are mounted to the shuttles. An additional ProcessSegment DrillBoard is defined that can be executed by the production system in order to drill a hole into the board. This process segment requires a drilling robot and the shuttle carrying the board needs to be located in a positioning unit that is within the reach of this robot. In our experiments, we have located the drilling robot next to the top right positioning unit, as is depicted in the lower part of Fig. 8. The results of the experiments are depicted using dashed lines. The task for the solver was to find a sequence of actions that would drill a hole in each of the boards.

Most importantly, the results show that it is feasible to convert ISA-95 models that include both inventory movements and manufacturing operations to PDDL and have it successfully solved. The new concepts (Predicates) and entities (Objects) required to formulate the new kind of operation have a significant impact on the solving performance. In fact, we could not compute a solution for the largest experiment (10:15) within our timeout frame of 300 s, which is why the diagram does not show values for this setting.

## VI. Conclusion

We have presented a conceptual mapping and a workflow for the transformation of ISA-95 models into a PDDL formalism in order to find sequences of production steps to fulfill certain manufacturing goals. We have successfully tested our approach in a use case where a chosen transportation system layout was tested whether it would fulfill certain logistics requirements.

To verify the proposed approach, we have developed a simple software tool that is able to execute the computed plan with real machinery. While the entire workflow has been tested and verified up to the point where real shuttles move in the aforementioned Industry 4.0 Testbed, we have focused on the conceptual model transformation part in this letter.

It is also worth noting that, while this letter has been focused on transportation systems, the generic approach and mapping strategy between ISA-95 and PDDL can be leveraged for other production-related problems as well. Briefly, we have already considered material manipulation (drilling) in one of our evaluation scenarios. Consequently, in a next step we would like to consider product assembly tasks in the production process. This should enable a flexible manufacturing system to create production plans for assembly-based lot-size 1 products automatically. What would be needed for such a scenario, would be the construction plan of the final product, as well as the consideration of machinery capabilities with respect to assembly operations.

The performance data presented in Section V is based on an non-optimized implementation: (i) the PDDL solver could be tweaked by experimenting with the parameters of its search algorithm. (ii) improvements could be achieved by parallelizing the tasks assigned to the PDDL solver. Currently, the 23 problems are solved sequentially in a simple for-loop—of course, the invocation of the solver could be done for several problems in parallel; most easily based on the number of cores the underlying platform provides. (iii) the encoding of the ISA-95 model in PDDL could be streamlined in a way that is more convenient to the solver (i.e., steps 1 and 2 of the core workflow could be improved). This argument has already been teased in Section III-B, where alternative PDDL encodings for specific ISA-95 constructs are mentioned.

Future work could also take into consideration more advanced versions of PDDL that would, e.g., enable the specification of *durative actions* [29], ultimately supporting parallelism of production tasks at the planning level. While such an approach could lead to finding highly efficient production plans, it might be too computationally expensive. Nevertheless, experiments in this direction seem worthwhile.